\documentclass[11pt, reqno, letter]{amsart}
\usepackage{setspace}
\usepackage[foot]{amsaddr}
\usepackage{tikz-qtree}
\usepackage{xcolor}
\usepackage{bm}
\usepackage{natbib}
\usepackage{url}

\newtheorem{proposition}{Proposition}
 
 \newcommand{\E}{\mathbb{E}}
 \newcommand{\pr}{\text{P}}
 \newcommand{\bX}{\textbf{X}}
 \newcommand{\bXij}{\textbf{X}_{ij}}

 \newcommand{\Yij}{Y_{ij}}
 \newcommand{\Tij}{T_{ij}}
 \newcommand{\yij}{y_{ij}}
 \newcommand{\tij}{t_{ij}}
 
 \newcommand{\hbeta}{\hat{\bm\beta}}
 
 \newcommand{\bbeta}{{\bm\beta}}
 \newcommand{\bgamma}{{\bm\gamma}}
 
\allowdisplaybreaks

\title[Accounting for recall bias in case-control studies]{Accounting for recall bias in case-control studies: a causal inference approach}
\author{Kwonsang Lee$^{1}$ and Francesca Dominici$^{2}$}
\address{$^{1}$ Department of Statistics, Sungkyunkwan University, Seoul, Republic of Korea}
\address{$^{2}$ Department of Biostatistics, Harvard T.H. Chan School of Public Health, Boston, MA, USA}

\begin{document}
 
\begin{abstract}
A case-control study is designed to help determine if an exposure is associated with an outcome. However, since case-control studies are retrospective, they are often subject to recall bias. Recall bias can occur when study subjects do not remember previous events accurately.  In this paper, we first define the estimand of interest: the causal odds ratio (COR) for a case-control study. Second, we develop estimation approaches for the COR and present estimates as a function of recall bias. Third, we define a new quantity called the \textit{R-factor}, which denotes the minimal amount of recall bias that leads to altering the initial conclusion. We show that a failure to account for recall bias can significantly bias estimation of the COR. Finally, we apply the proposed framework to a case-control study of the causal effect of childhood physical abuse on adulthood mental health.  
\end{abstract}


\maketitle
\setstretch{1.2}

\section{Introduction}
\label{sec1}

A case-control (or case-referent, case-noncase) study is designed to investigate whether there is evidence of an association between one or more risk factors and an outcome. In a case-control study, we first identify the cases (a group known to have the outcome) and the controls (a group known to be free of the outcome). Then, we look back in time to learn which subjects in each group had the exposure(s), comparing the frequency of the exposure in the case group to the control group. Therefore, by definition, a case-control study is always retrospective because it starts with an outcome and then collect retrospectively information about risks factors or exposures for the cases and the controls \citep{lewallen1998}. Compared to other study designs, case-control studies have several advantages. They are  time-efficient, inexpensive, and particularly suitable for rare outcomes \citep{mann2003observational}. However, they tend to be more susceptible to biases than other comparative studies \citep{schulz2002case}. 

The common problem in case-control studies is that it is difficult to measure exposure to risk factor correctly. Recall bias is a problem in studies that use self-reporting, such as  case-control studies and retrospective cohort studies. Recall bias is a systematic error that occurs when participants do not remember previous events/experience accurately or omit details: the accuracy and volume of memories may be influenced by subsequent events/experience. Recall bias can be random or differential \citep{rothman2012epidemiology}. Differential recall bias occurs when the exposure is under-reported for controls and over-reported for cases (or viceversa) \citep{barry1996differential, chouinard1995recall}. For example, in the data set analyzed in this paper there is evidence that adults with a mental health problem and high levels of anger scores (cases) tend to under-report their exposure, that is, whether or not they have been abused as children compared to the controls. This could be due to several factors, such as repression by the victim as a mean of self-protection or a desire to avoid discussing these experiences \citep{fergusson2000stability}. In addition to differential recall bias, random recall bias can often occur in case-control studies, but it equally occurs in case and control groups, for instance, due to memory failure itself. \cite{raphael1987recall}  pointed out that such random error will lead to measurement error that will usually lead to a loss of statistical power. On the contrary, differential recall bias is likely to lead to a biased estimate.

In this paper, we focus on differential recall bias rather than random bias. Many studies have documented that this differential recall bias can have a significant impact on associational measures \citep{chouinard1995recall, coughlin1990recall, drews1990impact, barry1996differential, greenland1996basic}. Despite of the importance of accounting for recall bias, there is yet a gap between previous studies about recall bias and current causal inference studies. First, to best of our knowledge, contributions regarding how recall bias affects measures with causal interpretations are scarce. For example, \cite{zhang2008estimating} proposed an approach based on a logistic regression model for estimating a marginal causal odds ratio (COR). \cite{Persson2013} proposed several approaches for estimating this marginal COR in case-control studies. However, neither accounted for recall bias. In contrast, many of the existing methods accounting for recall bias can reveal only an association between exposure and outcome. Furthermore, their analysis is limited to either estimating a marginal OR without adjusting for confounders or estimating a conditional OR relying on restricted models such as logistic regression. For instance, \cite{greenland1983correcting} proposed an approach for accounting for recall bias that uses a matrix correction in the context of a misclassification problem. However, the correcting matrix must be derived either from earlier studies or a validation study carried out on a subsample of the study subjects. Though using the matrix is conceptually simple, it is not feasible to adjust for confounders. Alternatively, \cite{barry1996differential} proposed a logistic regression method to assess the impact of recall bias on the conclusion by postulating a simpler misclassification model while adjusting for confounders. However, the confounder adjustment heavily depends on the logistic regression model, thus the target estimand is restricted to a conditional OR.  

The overall goal of this paper is to introduce a causal inference framework for case-control studies, for estimating the causal odds ratio (COR) in the presence of recall bias. More specifically, first, we define the CORs as a function of tuning parameters that quantify recall bias. The true COR and naive COR (which assumes that there is no recall bias even when recall bias is present in reality) can be analytically compared in terms of these tuning parameters. Second, we introduce two approaches for estimating COR as a function of these tuning parameters.  We focus on estimating the marginal COR that is often used to assess the exposure effect in the population as a whole. To estimate the marginal COR, we develop a maximum likelihood estimation method and a  stratification method based on the  prognostic score \citep{hansen2008prognostic}. Third, we also introduce and provide estimation approaches for other COR measures such as conditional and common CORs. Finally, we introduce the \textit{R-factor} defined as the  minimal amount of recall bias that would lead to a reporting of a spurious association between exposure and disease.    


\section{Notation and Odds Ratios in Case-control Studies}
\label{sec:notation}

\subsection{Causal Inference Framework in Case-Control Studies}

We start by introducing the notation for a matched case-control study that can cover the notation for an unmatched case-control study with one or more strata. In a matched  case-control study, first we stratify the population intro strata defined by age, gender, race or other demographics and then match the cases and controls within strata. \citep{breslow1982design}. Within each stratum, potential confounders that affect both exposure and outcome can be adjusted by certain criteria such as exact matching. If the outcome is rare, there are relatively few cases available in each stratum for the later analysis, which may result in providing inefficient estimates. Alternatively, researchers may want to find cases first and then match them to similar controls \citep{Breslow1980}. Age and sex are often exactly matched in this matched case-control design. 

Case-control studies are usually retrospective in nature. Most of the time, risk factors are retrospectively investigated to find the cause of the outcome, thus exposure to a risk factor is never randomized. A naive comparison of the prevalence of the outcome between the exposed and unexposed groups can be misleading due to confounding bias. To establish causation between exposure and outcome, we rely on the potential outcome framework \citep{neyman1990, rubin1974estimating}. Assume that there are $I$ strata. Each stratum $i$, $i=1, \ldots, I$, contains $n_i$ individuals. There are $N = \sum_{i=1}^{I} n_i$ individuals in total. We denote by  $ij$ the $j$th individual in stratum/matched set $i$ for $j=1, \ldots, n_i$. For example, if the data is collected from an unmatched case-control design without stratification, then there is only one stratum. We let $T_{ij}=1$,  to indicate that individual $ij$ was exposed to a certain risk factor, and  $T_{ij}=0$ otherwise. We can define potential outcomes as follows; if $T_{ij}=0$, individual $ij$ exhibits response $Y_{ij}(0)$, whereas if $T_{ij}=1$ then individual $ij$ exhibits $Y_{ij}(1)$. Depending on whether individual $ij$ was exposed or not, only one of the two potential outcomes can be observed. The response exhibited by individual $ij$ is $Y_{ij} = T_{ij} Y_{ij}(1) + (1-T_{ij})Y_{ij}(0)$. In this paper, both $Y_{ij}(1)$ and $Y_{ij}(0)$ are assumed to be binary. If $Y_{ij}=1$, individual $ij$ is considered as a case and if $Y_{ij}=0$ then individual $ij$ is a control. Let $\bXij$ denote a vector of covariates. In a matched case-control study with exact matching, within the same stratum $i$, two individuals $ij$ and $ij'$ share the same covariates (i.e., $X_{ij} = X_{ij'}$). For an unmatched case-control study, a function of $X_{ij}$ such as the propensity score \citep{rosenbaum1983central} can be used to construct strata adjusting for confounding bias. In this case, two individual $ij$ and $ij'$ may have different values of the covariates, but the same value of the propensity score.

\subsection{Definition and Identification of Conditional and Marginal Causal Odds Ratios}
\label{ss:def_odds_ratio}
In this section, we introduce the parameter of interest, which is \textit{marginal causal odds ratio (COR)} defined as $\psi = \{p_1 ( 1- p_0)\}/\{p_0 (1 - p_1)\}$ where $p_t = \E[\Yij(t)]; t=0,1$. In some instances we might be interested in the \textit{conditional COR} at a given level of $\bX = x$,  which is defined as $\psi(x) = \{p_1(x) (1-p_0(x))\}/\{p_0(x) (1 - p_1(x))\}$ where $p_t(x) = \E[\Yij(t) \mid \bXij = x]$. We note that $p_t = \E_{\bX}[p_t(x)]$. However, in general $\psi \neq \E_{\bX}[\psi(x)]$, which is often referred as the non-collapsibility of the odds ratios \citep{greenland1986identifiability}. 

To identify these causal parameters, we consider two assumptions: (1) unconfoundedness and (2) positivity. The unconfoundedness assumption means that the potential outcomes $(\Yij(1), \Yij(0))$ are conditionally independent of the treatment $\Tij$ given $\bXij $, i.e., $(\Yij(1), \Yij(0)) \perp\!\!\!\perp \Tij \>|\> \bXij$. The second assumption means that the probability $\pr(\Tij = 1 \mid \bXij)$ lies in $(0,1)$. These assumptions together are often called \textit{strong ignorability}. We introduce $p_{y|t}(x) = \pr(\Yij = y \mid \Tij = t, \bXij = x)$ for $t=0, 1$ and $y=0, 1$. We note that these probabilities can be computed by using observable variables. Under the strong ignorability assumption, $p_0(x), p_1(x)$ based on potential outcomes can be identified as $p_{1|0}(x), p_{1|1}(x)$ respectively. Also  the conditional and marginal CORs can be identified as
\begin{equation}
	\psi(x) = \frac{p_{1|1}(x) p_{0|0}(x)}{p_{1|0}(x) p_{0|1}(x)}, \quad \psi = \frac{\E[p_{1|1}(x)] \E[p_{0|0}(x)]}{\E[p_{1|0}(x)] \E[p_{0|1}(x)]}, 
	\label{eqn:or_identification}
\end{equation}
Using the identification results, many estimation methods based on the propensity score have been proposed. The performance of methods for the marginal and conditional ORs are compared in \cite{austin2007performance} and \cite{Austin2007a} respectively. However, most of the developed methods are not suitable in the presence of recall bias.

\section{Impact of Recall Bias on Causal Parameters}
\label{sec:recall}

\subsection{Recall Bias Model} 

In this paper, we consider situations with differential recall bias where the exposure is over-reported (under-reported)  differently among cases and controls. In presence of recall bias, the underlying true exposure $T_{ij}$ is not observed. Instead, we observe the biased exposure $T_{ij}^*$. If there is no recall bias, then $\Tij = \Tij^*$. 

We introduce two tuning parameters to measure differential over-reporting recall bias:
\begin{align}
	\eta_1 &= \pr(T_{ij}^*=1 \mid Y_{ij}=1, T_{ij}=0, \bXij = x) \nonumber\\
	\eta_0 &= \pr(T_{ij}^*=1 \mid Y_{ij}=0, T_{ij}=0, \bXij = x)
	\label{model:recall_bias}
\end{align}
The parameters $(\eta_1,\eta_0)$ represent the probability of over-reporting among cases and the controls respectively. We implicitly assume in \eqref{model:recall_bias} that the magnitude of recall bias does not depend on covariates $\bXij$. Also, for each $ij$, $T_{ij}=1$ implies $T_{ij}^* =1$, but $T_{ij}=0$ implies either $T_{ij}^*=0$ or $T_{ij}^*=1$. Therefore, recall bias occurs only when $\Tij=0$. If $T_{ij}=0$, $\Yij(0)$ is observed as $\Yij$, and the observed exposure $T_{ij}^*$ depends on $Y_{ij}(0)$. Similarly, we can define another set of two tuning parameters in under-reporting situations,
\begin{align}
	\zeta_1  & =  \pr(\Tij^* = 0 \mid \Yij(1) = 1, \Tij = 1, \bXij = x) \nonumber \\
	\zeta_0 &=  \pr(\Tij^* = 0 \mid \Yij =0, \Tij = 1, \bXij = x)
	\label{model:recall_bias_under}
\end{align}
The parameters $(\zeta_0,\zeta_1)$ represent the probability of under-reporting among cases and the controls respectively. We use only one of the two sets of the parameters depending on the situation whether exposure is over-reported or under-reported. Note that if there is no recall bias, then $\eta_1 =\eta_0=0$ (or $\zeta_1 =\zeta_0=0$).

For our example that will be discussed in Section~\ref{s:example}, we consider a study of child abuse and adult mental health, especially anger. Adults are likely not to report their childhood abuse if there are any. Thus, exposure was generally under-reported. We use the parameter set $(\zeta_0, \zeta_1)$. The parameter $\zeta_1$ ($\zeta_0$) is the proportion of adults with higher (lower) anger scores that fail to recall child abuse correctly. In this study, it is known that under-reporting child abuse is equally likely for cases and controls. This information obtained from previous literature can impose an additional restriction such as $\zeta_1 = \zeta_0$. We will discuss two analyses with and without this additional information in Section~\ref{s:example}.

\subsection{Change in Target Parameters Due to Recall bias }

In this section, we investigate the impact of recall bias on the marginal and conditional CORs, analytically. If there is no recall bias and the exposure $\Tij$ is observed correctly, the conditional and marginal CORs, $\psi(x)$ and $\psi$, can be identified based on the conditional probabilities $p_{y|t}(x)$ as in \eqref{eqn:or_identification}. In presence of recall bias, if we make inference based on $\Tij^*$ rather than $\Tij$, then we obtain a biased estimate of the COR since $(\Yij(1), \Yij(0)) \not\perp\!\!\!\perp \Tij^* \mid \bXij$. To describe this, consider the probabilities based on observable variables, $p_{y|t}^*(x) = \pr(\Yij = y \mid \Tij^* = t, \bXij = x)$ for $y=0,1$ and $t=0,1$. In the presence of recall bias, any consistent estimator of $\psi$  target a different estimand $\psi^*$ (or $\psi^*(x)$), 
\begin{equation*}
	\psi^*(x) = \frac{p_{1|1}^*(x) p_{0|0}^*(x)}{p_{1|0}^*(x) p_{0|1}^*(x)}, \quad \psi^* =  \frac{\E[p_{1|1}^*(x)] \E[p_{0|0}^*(x)]}{\E[p_{1|0}^*(x)] \E[p_{0|1}^*(x)]}. 
\end{equation*}
The following propositions compare $\psi(x), \psi$ with $\psi^*(x), \psi^*$ respectively.
\begin{proposition}
	For given $0 \leq \eta_0, \eta_1  \leq 1$, 
	$$
	\psi(x) \leq \psi^*(x) \Leftrightarrow {q}_1^*(x) \eta_0 \leq {q}_0^*(x) \eta_1 \quad \text{for each } x
	$$
	where $q_y^*(x) = \pr(\Tij^*=1 | \Yij=y, \bXij=x)$ for $y=0, 1$. Similarly, if the source of bias is under-reporting of exposure, for given $0 \leq \zeta_0, \zeta_1 \leq 1$, then 
	$$
	\psi(x) \leq \psi^*(x) \Leftrightarrow \{1- q_1^*(x)\} \zeta_0 \geq \{1-q_0^*(x)\} \zeta_1.
	$$
	\label{prop1}
\end{proposition}
This proposition implies that the true conditional COR $\psi(x)$ is always smaller than $\psi^*(x)$ if ${q}_1^*(x) \eta_0 \leq {q}_0^*(x) \eta_1$. Using models or stratification for $q_y^*(x)$, the condition can be checked for fixed values of $(\eta_0, \eta_1)$ or $(\zeta_0, \zeta_1)$. In particular, if $\eta_0 =0$ (i.e., no over-reporting bias for controls), $\psi(x) \leq \psi^*(x)$.  Proposition~\ref{prop1} shows the condition for conditional CORs $\psi(x) \leq \psi^*(x)$, but this condition cannot be generalized to the marginal COR case due to non-collapsibility \citep{greenland1999confounding}. Under certain additional conditions, $\psi \leq \psi^*$ can be claimed.

\begin{proposition}
	Assume that exposure is over-reported.
	\begin{enumerate}
		\item[(i)] Suppose $\eta_0 \leq \eta_1$. If $\psi(x) \leq \eta_1 / \eta_0$, then $\psi(x) \leq \psi^*(x) $. Furthermore, if $\max_{x} \psi(x) \leq \eta_1 / \eta_0$, then $\psi \leq \psi^* $.
		
		\item[(ii)] Suppose $\eta_0 \geq \eta_1$. If $\psi(x) \geq \eta_1 / \eta_0$, then $\psi(x) \geq \psi^*(x)$. Furthermore, if $\min_{x} \psi(x) \geq \eta_1 / \eta_0$, then $\psi \geq \psi^*$. 
	\end{enumerate}
	
	\label{prop2}
\end{proposition} 
This proposition means that, when exposure is over-reported, if the maximum of the conditional COR, $\max_{x} \psi(x)$, is less than or equal to $\eta_1/\eta_0$, then recall bias leads to $\psi \leq \psi^*$. Also, if $\eta_0=0$, then $\psi(x)$ is always less than $\eta_1/\eta_0$ and thus $\psi(x) \leq \psi^*(x)$ for all $x$ and thus $\psi \leq \psi^*$. When $\eta_0 = \eta_1$, $\psi(x) \leq \psi^*(x)$ if $\psi(x) \leq 1$. A similar argument can be applied to the situation where exposure is under-reported. When assuming that exposure is under-reported, the following statements hold: (i) Suppose $\zeta_0 \leq \zeta_1$. If $\psi(x) \geq \zeta_0 / \zeta_1$, then $\psi(x) \geq \psi^*(x) $. Furthermore, if $\min_{x} \psi(x) \geq \zeta_0 / \zeta_1$, then $\psi \geq \psi^* $, and (ii) Suppose $\zeta_0 \geq \zeta_1$. If $\psi(x) \leq \zeta_0 / \zeta_1$, then $\psi(x) \leq \psi^*(x)$. Furthermore, if $\max_{x} \psi(x) \leq \zeta_0 / \zeta_1$, then $\psi \leq \psi^*$. Similarly, when $\zeta_0 = \zeta_1$, $\psi(x) \geq \psi^*(x)$ if $\psi(x) \geq 1$. Proofs for both propositions are in the Supplementary Materials. 

Propositions~\ref{prop1} and \ref{prop2} enable us to assess the impact of recall bias on both marginal and conditional CORs before starting the main analysis. For example, it is known from previous literature that child abuse and adult anger are positively associated based on logistic regression, which implies $\psi(x) \geq 1$ for each $x$. Given the additional information of $\zeta_0 = \zeta_1$, from Proposition~\ref{prop2} (i) for under-reported exposures, we expect $\psi(x) \geq \psi^*(x)$, and further $\psi  \geq \psi^*$. This relationship shows that if the underlying true COR $\psi(x)$ is recovered by accounting for recall bias, then it should be greater than the biased estimand $\psi^*(x)$. Even without the prior knowledge, Proposition~\ref{prop1} provides a way to predict the impact of recall bias. Using a model for $q_y^*(x), y=0,1$, it is possible to check whether the condition $\{1- q_1^*(x)\} \zeta_0 \geq \{1-q_0^*(x)\} \zeta_1$ is satisfied or not. For instance, a logistic regression model $q_y^*(x) = \exp(\beta_y Y + \beta_{\bX} \bX)/\{1 + \exp(\beta_y Y + \beta_{\bX} \bX)\}$ can be considered, and then used for assessing the condition with various values of $(\zeta_0, \zeta_1)$. Especially when using $\zeta_0 = \zeta_1$, the condition for $\psi(x) \geq \psi^*(x)$ is further simplified as $\beta_y \geq 0$.

\subsection{R-factor: how much recall bias is needed to alter the initial conclusion?}
\label{ss:rfactor}

\begin{figure}
	\centering
	\includegraphics[width=120mm]{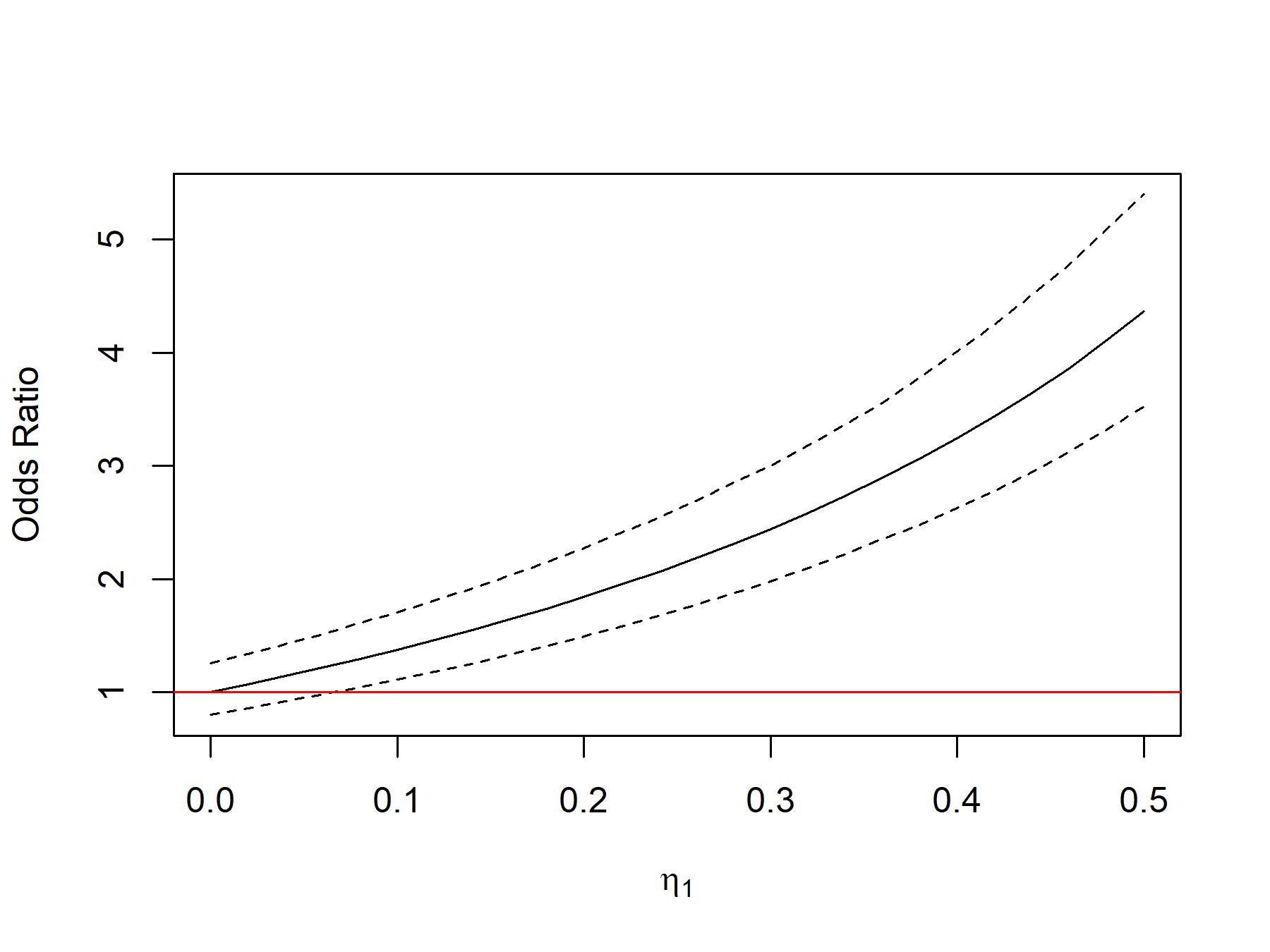}
	\caption{Changes in $\psi$ depending on the size of recall bias. Estimates and  95\% confidence intervals of ${\psi}$ are shown for different values of $\eta_1$. The red line represents the true COR $\psi=1$. }
	\label{fig:eff_recall_bias}
\end{figure}

The above propositions show that, when there is recall bias, estimators without controlling for recall bias can introduce bias. In this section, we illustrate that even a small amount of recall bias can bias the COR estimate significantly. To illustrate this, with a small simulation study. We assume take $N=2000$, just one $I=1$, and no confounding. We assume $T_{1j} \sim \text{Bernoulli}(0.3)$, $Y_{1j}(1) \sim \text{Bernoulli}(0.25)$ and $Y_{1j}(0) \sim \text{Bernoulli}(0.25)$. The marginal COR $\psi$ is 1. Then, we compute $Y_{1j} = Y_{1j}(1)T_{1j} + Y_{1j}(0)(1-T_{1j})$. We consider a situation of over-reporting among the cases only, that is $\eta_1>0$ and $\eta_0=0$. Therefore, if $T_{1j}=1$, then $T_{1j}^*=1$, but if $T_{1j}=0$, then $T_{1j}^* \sim \text{Bernoulli}(\eta_1)$. For each simulated dataset, we calculate the bias of $\hat{\psi}^*$  with respect to the true $\hat{\psi}$ as a function of $\eta_1$. Figure~\ref{fig:eff_recall_bias} shows the estimate $\hat{\psi}^*$ and its $95\%$ confidence intervals (CI) as a function of $\eta_1 \in [0, 0.5]$. As $\eta_1$ increases, $\hat{\psi}^*$ increases rapidly. We can see that failing to account for recall bias can lead to a misleading conclusion even for a small amount of recall bias. 

In this paper, we introduce a \textit{R-factor} to quantify the minimum amount of recall bias that can alter the initial conclusion. As shown in Figure~\ref{fig:eff_recall_bias}, we emphasize that, for $\eta_1 > 0.06$, we would conclude that there is a statistically significant relationship between exposure and disease when in reality there is none, $\psi=1$. In this context, we define the R-factor is $6\%$, which means that if among the cases at least $6\%$  reported to be exposed where in reality they were not, the initial conclusion of no effect would be altered. More specifically for $\eta_1 > 0.06$, since the 95\% CI does not contain 1, we would reject the null hypothesis even though the true COR is 1.  

The R-factor can be differently defined in a different context. For instance, suppose the initial conclusion that rejects the null hypothesis of no effect. Then, the R-factor can be defined as the minimum amount of recall bias (in this case, in terms of $\eta_0$ while fixing $\eta_1=0$) to make the same hypothesis not to be rejected. The R-factor can be used for a summary statistic that shows the sensitivity of the conclusion. 

\section{Two Methods for Recovering the Marginal Causal Odds Ratio in the Presence of Recall Bias}
\label{sec:method}

In this section, we propose two estimation methods that provide a consistent estimate of the marginal COR $\psi$ in presence of recall bias and confounding: (1) maximum likelihood estimation and (2) stratification using prognostic scores. For given values of $(\eta_0, \eta_1)$ or $(\zeta_0, \zeta_1)$, to get a consistent estimate of $\psi$, the first ML-based method requires the models for two potential outcomes and exposure to be correctly specified. The stratification-based method requires less model assumptions. Stratification can be implemented based on either propensity scores or prognostic scores \citep{hansen2008prognostic}. In the following subsections, we discuss these two estimation methods in more detail. For simplicity, we consider the over-reported exposure case with the tuning parameters $(\eta_0, \eta_1)$. Under-reported exposure with  $(\zeta_0, \zeta_1)$ can be dealt with in a similar way. 

\subsection{Maximum Likelihood Estimation (ML)}
\label{ss:mle}

Consider the outcome models $m_t(x) = \pr(\Yij = 1 | \Tij = t, \bXij = x), t=0, 1$ that are the models for the two probabilities, $p_{1|1}(x)$ and $p_{1|0}(x)$. Also, consider the model $e(x)$  for the probability $\pi(x) = \pr(\Tij=1 | \bXij = x)$ that is often called the \textit{propensity score} \citep{rosenbaum1983central}. In Section~\ref{ss:def_odds_ratio}, we discussed that the probability $p_1(x)$ can be identified as $p_{1|1}(x)$, thus can be estimated by $m_1(x)$. In the absence of recall bias, either $m_t(x)$ or $e(x)$ is required to be correctly specified to obtain a consistent estimate. However, in presence of recall bias, since we do not observe $\Tij$,  $m_t(x)$ nor $e(x)$ cannot be estimated from the observable dataset. We can estimate the marginal COR as a function of the tuning parameters of the recall bias model.

The first method presented in this subsection uses maximum likelihood estimation. To construct the likelihood function, it is required to specify both $m_t(x)$ and $e(x)$ to obtain an estimate for given values of $(\eta_0, \eta_1)$. Using the recall bias model~\eqref{model:recall_bias}, the joint probability $\pr(\Yij, \Tij^* |  \bXij)$ of observable variables can be represented by a function of $m_0(x), m_1(x)$ and $e(x)$. We assume models $m_t(x; \gamma_t), t=0, 1$ and $e(x; \beta)$ with parameters $\gamma_t$ and $\beta$. For instance, logistic regressions can be used such as $m(T, \bX; \bgamma) = \exp(\gamma_t T + \gamma_{\bX}^T \bX)/(1+ \exp(\gamma_t T + \gamma_{\bX}^T \bX ))$ with $m_1(\bX) = m(1, \bX; \bgamma)$ and $m_0(\bX) = m(0, \bX; \bgamma)$ and $e(\bX) = \exp(\bbeta^T \bX) /(1 + \exp(\bbeta^T \bX))$. These model parameters can be estimated by solving the following maximization problem, 
\begin{equation}
	\hat{\bm\theta} = (\hat{\beta}, \hat{\gamma}_0, \hat{\gamma}_1) = \arg \max_{\beta, \gamma_0, \gamma_1} \sum_{i=1}^{I} \sum_{j=1}^{n_i} \log \pr(\Yij = \yij, \Tij^* = \tij^* | \bXij = x),
	\label{eqn:mle}
\end{equation}
where
\begin{align*}
	\pr(\Yij=1, \Tij^* = 1 | \bXij =x) &= m_1(x; \gamma_1) e(x; \beta) + \eta_1 m_0(x; \gamma_0)\{1-e(x; \beta)\}\nonumber \\
	\pr(\Yij=0, \Tij^*=1 | \bXij = x) &= \{1-m_1(x; \gamma_1)\} e(x; \beta) + \eta_0 \{1-m_0(x; \gamma_0)\} \{1-e(x; \beta)\}\nonumber \\
	\pr(\Yij=1, \Tij^*=0 | \bXij = x) &= (1-\eta_1)m_0(x; \gamma_0) \{1-e(x; \beta)\}\nonumber \\
	\pr(\Yij=0, \Tij^*=0 | \bXij = x) &= (1-\eta_0)\{1-m_0(x; \gamma_0)\} \{1-e(x; \beta)\}.
	\label{model:mle_true}
\end{align*}
Once we obtain the estimate $\hat{\bm \theta}$,  we can compute $\hat{m}_t(x) = m_t(x; \hat{\gamma}_t)$ and $\hat{e}(x) = e(x; \hat{\beta})$. The marginal probability ${p}_y, y=0, 1$ is then estimated by taking sample averages of $\hat{m}_t(x)$,
$$
\hat{p}_{1, ML} = \frac{1}{N} \sum_{i=1}^{I} \sum_{j=1}^{n_i} \hat{m}_1(\bXij), \quad \hat{p}_{0, ML} = \frac{1}{N} \sum_{i=1}^{I} \sum_{j=1}^{n_i} \hat{m}_0(\bXij) 
$$
The marginal COR can be estimated by 
$$
\hat{\psi}_{ML} = \frac{\hat{p}_{1, ML} ( 1- \hat{p}_{0, ML})}{\hat{p}_{0, ML} (1- \hat{p}_{1, ML})}.
$$
Since the estimate $\hat{\bm \theta}$ can vary with the values of $\eta_0$ and $\eta_1$, the estimator $\hat{\psi}_{ML}$ can be considered as a function of $\eta_0$ and $\eta_1$. We will use notation $\hat{\psi}_{ML}(\eta_0, \eta_1)$ if an explicit expression is necessary. Unlike usual situations in causal inference, to obtain a valid estimate of $\hat{\psi}_{ML}$, both $m_t(x; \gamma_t)$ and $e(x; \beta)$ have to be estimated correctly. However, when obtaining $\hat{\psi}$, the estimated parameter $\hbeta$ is not used. This parameter can be understood as a nuisance parameter. We note that since $\Tij$ is not observable, the inverse probability weighted estimator as proposed in \cite{forbes2008inverse} cannot be considered here. 

Under the recall bias model~\eqref{model:recall_bias}, if $(\eta_0, \eta_1)$ are known and the three models for $m_0, m_1, e$ are correctly specified, then $\hat{\psi}$ is a consistent estimator for $\psi$. For different values of $\eta_1$ and $\eta_0$, $\hat{\psi}$ can be changed but $\hat{\psi}^*$ remains unchanged. Therefore, we can consider $\hat{\psi}$ as an estimator that recovers the true marginal COR when we believe $\eta_0$ and $\eta_1$ are correctly specified. However, $\eta_0$ and $\eta_1$ are unknown in practice. For data applications, it is recommended to consider a plausible region of $(\eta_0, \eta_1)$ based on previous knowledge. Then, the causal conclusion can be evaluated with the considered region.

Also, for given $(\eta_0, \eta_1)$, the asymptotic variance of the estimator $\hat{\psi}(\eta_0, \eta_1)$ can be obtained by using the theory of M-estimation \citep{stefanski2002calculus}. However, it is not recommended because the formula for the variance is difficult to derive in general. Instead, the bootstrap procedure can be considered to construct confidence intervals for $\psi$.

\subsection{Stratification}
\label{ss:stratification}

\begin{table}
	\centering
	\caption{The $2 \times 2$ contingency observed table for the $i$th stratum}
	\begin{tabular}{c|cc|c}
		& Case & Control & \\
		\hline
		Exposed $(T^*=1)$ & $a_i^*$ & $b_i^*$ & $a_i^*+b_i^*$ \\
		Not Exposed $(T^*= 0)$ & $c_i^*$ & $d_i^*$ & $c_i^*+d_i^*$ \\
		\hline
		& $a_i^* + c_i^*$ & $b_i^*+d_i^*$ & $n_i^*$
	\end{tabular}
	\label{tab:contingency}
\end{table}

Stratification can be alternatively used to estimate $\psi$ by aiming to balance the covariate distributions between exposed and unexposed groups. Among stratification based methods, \cite{graf2008comments} proposed an estimator for $\psi$ based on stratum-specific probabilities for strata defined by the propensity score, $p_{1i} = \E_{\bX | \textbf{stratum i}}[p_1(x)]$ and $p_{0i} = \E_{\bX | \textbf{stratum i}}[p_0(x)]$. If the assumption that $(\Yij(1), \Yij(0))$ is independent of $\Tij$ within each stratum $i$ holds, then these probabilities can be identified from the $2 \times 2$ table generated by stratum $i$. However, due to recall bias, $\Tij^*$ is observed instead of $\Tij$. Under the model~\eqref{model:recall_bias}, for $p_{yt}(x) = \E(\Yij = y, \Tij = t \mid \bXij = x)$ and $p_{yt}^* (x) = \E(\Yij = y, \Tij^* = t \mid \bXij = x)$, the following relationships hold, 
\begin{align*}
	p_{11}(x) = p_{11}^*(x) - \frac{\eta_1 }{1-\eta_1} p_{10}^*(x), &\quad p_{10}(x) = \frac{p_{10}^*(x)}{1-\eta_1} \\
	p_{01}(x) = p_{01}^*(x) - \frac{\eta_0 }{1-\eta_0} p_{00}^*(x), &\quad p_{00}(x) = \frac{p_{00}^*(x)}{1-\eta_0}.
\end{align*}
For each generated stratum, assume that Table~\ref{tab:contingency} is observed. Using the above relationships, the probabilities $p_{1i}$ and $p_{0i}$ are estimated by $\hat{p}_{1i} = a_i/\{a_i + b_i\}, \hat{p}_{0i} = c_i/\{c_i+d_i\}$
where $a_i = \{a_i^* - \eta_1 (a_i^* + c_i^*)\}/(1-\eta_1)$, $b_i = \{b_i^* - \eta_0 (b_i^* + d_i^*)\}/(1 - \eta_0)$, $c_i = c_i^*/(1-\eta_1)$ and $d_i = d_i^*/(1-\eta_0)$. Similarly, for the under-reporting exposure case, $a_i = a_i^*/(1-\zeta_1)$, $b_i = b_i^*/(1-\zeta_0)$, $c_i = \{c_i^* - \zeta_1(a_i^* + c_i^*)\}/(1-\zeta_1)$, and $d_i = \{d_i^* - \zeta_0(b_i^* + d_i^*)\}/(1-\zeta_0)$. The marginal probabilities can be estimated by weighted average of these stratum-specific probabilities with weights $s_i = n_i/N$, i.e., $\hat{p}_{1, S} = \sum_{i=1}^{I} s_i \hat{p}_{1i}$ and $\hat{p}_{0, S} = \sum_{i=1}^{I} s_i \hat{p}_{0i}$. Therefore, the marginal COR is estimated by 
$$
\hat{\psi}_{S} = \frac{\hat{p}_{1, S} (1-\hat{p}_{0, S})}{\hat{p}_{0, S} (1-\hat{p}_{1, S})}.
$$
The estimator $\hat{\psi}_{S}$ was discussed and the variance formula of $\hat{\psi}_{S}$ was given in \cite{stampf2010estimators}. We provide a modified variance formula that account for recall bias in the Supplementary Materials. However, this variance estimator is valid when strata are fixed. If strata are decided by the observed data, bootstrap procedures are recommended. 

Compared to the estimator $\hat{\psi}_{ML}$, this approach does not use outcome logistic regressions ($m_0$ and $m_1$). Therefore, $\hat{\psi}_{S}$ require less modeling assumptions. Also, to obtain $\hat{\psi}_{S}$, less computation effort is required. However, as many of stratification-based methods, this method relies on the assumption that stratification achieves covariate balance at least approximately. Furthermore, strata are formed based on the biasedly estimated propensity score $\hat{e}^*(x)$ using $\Tij^*$ instead of unobservable $\Tij$. It is not feasible to compare the covariate distributions between the exposed ($T_{ij}=1$) and unexposed ($T_{ij}=0$) groups. Since we assume that recall bias is independent of covariates conditioning on observed outcome, if $\eta_0 = \eta_1$, the covariate balance between the $\Tij^*=1$ and $\Tij^*=0$ groups is asymptotically same as that between the $\Tij=1$ and $\Tij=0$ groups. 

As we discussed, constructing strata based on the propensity score can be problematic if $\eta_1$ and $\eta_0$ are significantly different from 0. Instead of using the propensity score, the prognostic score \citep{hansen2008prognostic} can be used to construct strata. If there is $\Psi(\bXij)$ such that $ \Yij(0) \perp\!\!\!\perp \bXij \>|\> \Psi(\bXij)$, we call $\Psi(x)$ the prognostic score. Like propensity score stratification, prognostic stratification permits estimation of exposure effects within the exposed group. If it is assumed that there is no effect modification, prognostic stratification is valid for estimating overall exposure effects. For instance, if  $m(\Tij, \bXij; \bgamma) = \exp(\gamma_t \Tij + \gamma_{\bX}^T \bXij)/(1+ \exp(\gamma_t T + \gamma_{\bX}^T \bXij ))$ is assumed, $\Psi(\bXij) = \gamma_{\bX}^T \bXij$ is the prognostic score. As with propensity scores, stratification on the prognostic score leads to a desirable and balanced structure. Since we do not know $\Psi(\bXij)$ a priori, this has to be estimated from the data. Since exposure was over-reported, we know $\Tij^* = 0$ always implies $\Tij = 0$. We can estimate $\gamma_{\bX}^T$ from using the data of the $\Tij^*=0$ group, and estimate $\Psi(\bXij)$ for all individuals. 

Furthermore, the stratification method provides an explicit expression of the estimator in terms of stratum-specific cell counts. Thus, the stratification estimate is tractable. In particular, the estimator $\hat{\psi}_{S}(\eta_0, \eta_1)$ is an increasing function of $\eta_0$ and a decreasing function of $\eta_1$. 
\begin{proposition}
	The function $\hat{\psi}_{S}(\eta_0, \eta_1)$ satisfies
	\begin{align*}
		\hat{\psi}_{S}(\eta_0, \eta_1) \leq \hat{\psi}_{S}(\eta_0', \eta_1) &\>\text{ if } \eta_0 \leq \eta_0' \\
		\hat{\psi}_{S}(\eta_0, \eta_1) \geq \hat{\psi}_{S}(\eta_0, \eta_1') &\>\text{ if } \eta_1 \leq \eta_1'
	\end{align*}
	\label{prop3}
\end{proposition}
The proof of this proposition is given in the Supplementary Materials. The results of this proposition is straightforward. For example, if it is suspected that exposure is more over-reported among controls (i.e., a higher $\eta_0$), then the odds of exposure among controls is thought to be more inflated. Therefore, the recovered COR value is higher. Similarly, we can prove that $\hat{\psi}_{S}(\zeta_0, \zeta_1)$ is a decreasing function of $\zeta_0$ and an increasing function of $\zeta_1$.

\section{Simulation}
\label{sec:simulation}

We conduct a simulation study for comparing the performance of the methods we proposed in the previous sections - (1) ML method and (2) Stratification method. We consider two approaches for the Stratification (S) method: one based on the propensity score stratification and the other based on the prognostic score stratification. We call the former S$_{prop}$ and the latter S$_{prog}$. 

We consider four binary covariates, say $X_{i1}, X_{i2}, X_{i3}, X_{i4}$. All of the covariates are finely balanced in the sample population, which means $2^4 = 16$ possible combinations have $n/16$ individuals each. For each individual $i$, we randomly generate an exposure and potential outcomes:
$$
T_i \sim \text{Bernoulli}(p_{i,T}), Y_i(0) \sim \text{Bernoulli}(p_{i, Y(0)}), Y_i(1) \sim \text{Bernoulli}(p_{i, Y(1)})
$$
where 
\begin{align*}
	\text{logit}(p_{i,T}) &= \beta_0 + \beta_1 X_{i1} + \beta_2 X_{i2} + \beta_3 X_{i3} + \beta_4 X_{i4} + \beta_{1,2} X_{i1}X_{i2}\\
	\text{logit}(p_{i,Y(0)}) &= \gamma_0 + \gamma_1 X_{i1} + \gamma_2 X_{i2} + \gamma_3 X_{i3} + \gamma_4 X_{i4} + \gamma_{1,2} X_{i1}X_{i2} \\
	\text{logit}(p_{i,Y(1)}) &= \gamma_T + \text{logit}(p_{i,Y(0)}).
\end{align*}
Since we do not consider effect modification, the effect of an exposure on an outcome is constant for every individual, and the true conditional COR is $\exp(\gamma_T)$. However, the true marginal COR is computed by $\{\bar{p}_{Y(1)}(1-\bar{p}_{Y(0)})\}/\{\bar{p}_{Y(0)}(1-\bar{p}_{Y(1)})\}$ where $\bar{p}_{Y(0)} = (1/n) \sum_{i=1}^{n} p_{i,Y(0)}$ and $\bar{p}_{Y(1)} = (1/n) \sum_{i=1}^{n} p_{i,Y(1)}$. 

Due to recall bias, we cannot observe an exposure $T_i$, instead we observe the biased exposure $T_i^*$. For this simulation study, we assume that exposure is over-reported. We generate $T_i^*$ based on observed outcome $Y_i = Y_i(1) T_i + Y_i(0) (1-T_i)$ such as
$$
T_i^* = T_i + (1-T_i)Y_i \cdot RB_{i1} + (1-T_i)(1-Y_i) \cdot RB_{i0}$$
where $RB_{i1} \sim \text{Bernoulli}(\eta_1)$ and $RB_{i0} \sim \text{Bernoulli}(\eta_0)$. $RB_{i1}$ and $RB_{i0}$ are indicators of recall bias under $Y_i=1$ and $Y_i=0$ respectively. If $T_i = 1$, $RB_{i1}$ and $RB_{i0}$ have no impact on $T_i^*$ since $T_i^*$ always 1. 

We consider logistic regression models for exposure and outcome only with the four covariates, but without considering the interaction term. Thus, if $\beta_{1,2} \neq 0$, the exposure model is misspecified. Similarly, if $\gamma_{1,2} \neq 0$, then the outcome model is misspecified. We consider four simulation scenarios where the exposure and outcome models are  correctly specified or misspecified: (i) (cor, cor), (ii) (mis, cor), (iii) (cor, mis), and (iv) (mis, mis). For example, (mis, cor) means the exposure model is misspecified (i.e., $\beta_{1,2} \neq 0$), but the outcome model is correctly specified (i.e., $\gamma_{1,2} = 0$). For our simulation study, we set $(\beta_0, \beta_1, \beta_2, \beta_3, \beta_4) = (-1, 1, -1, 1, 0)$ and $(\gamma_0, \gamma_1, \gamma_2, \gamma_3, \gamma_4) = (-2, 2, -2, 0, 1)$. For each simulation scenario, we set $(\beta_{1,2}, \gamma_{1,2}) = (0, 0)$ for (cor, cor), $(2, 0)$ for (mis, cor), $(0, -2)$ for (cor, mis), and $(2, -2)$ for (mis, mis). We compare the considered methods in terms of how they can successfully recover the true marginal CORs under different model misspecification scenarios.

\begin{table}
	\centering
	\caption{Performance of the estimation methods for recovering the marginal COR. Three methods are compared, (1) maximum likelihood, (2) stratification based on propensity scores, and (3) stratification based on prognostic scores. The Crude method that does not account for recall bias is also reported. The log values of the marginal CORs are reported. The size of recall bias is $(\eta_0, \eta_1) = (0.1, 0.1)$.}
	\begin{tabular}{lr|rrrrr}
		\hline
		& & \multicolumn{5}{c}{Marginal OR} \\
		Scenario & $n$ & True & Crude & ML & S$_{prop}$ & S$_{prog}$ \\
		\hline
		(cor, cor) & 800 & 0.000 & 0.597 &  0.002 & 0.102 & 0.046 \\
		& & 0.357 & 0.917 &  0.358 & 0.466 & 0.401 \\
		& & 0.706 & 1.231 &  0.714 & 0.827 & 0.751 \\[0.1cm]
		& 2000 & 0.000 & 0.591 & -0.001 & 0.115 & 0.040\\
		& & 0.357 & 0.919 &  0.360 & 0.477 & 0.400 \\
		& & 0.706 & 1.226 &  0.704 & 0.827 & 0.740 \\
		\hline
		(mis, cor) & 800 & 0.000 & 0.481 &  0.008 & -0.065 &  0.016 \\
		& & 0.357 & 0.826 &  0.367 &  0.311 &  0.371 \\
		& & 0.706 & 1.174 &  0.731 &  0.690 &  0.729 \\[0.1cm]
		& 2000 & 0.000 & 0.472 & -0.002 & -0.064 & -0.001 \\
		& & 0.357 & 0.822 &  0.364 &  0.317 &  0.360 \\
		& & 0.706 & 1.167 &  0.720 &  0.686 &  0.709 \\
		\hline
		(cor, mis) & 800 & 0.000 & 0.681 & 0.010 & 0.110 & 0.054 \\
		& & 0.310 & 0.958 & 0.313 & 0.428 & 0.364 \\
		& & 0.607 & 1.229 & 0.603 & 0.726 & 0.660 \\[0.1cm]
		& 2000 & 0.000 & 0.674 & 0.005 & 0.123 & 0.049 \\
		& & 0.310 & 0.955 & 0.307 & 0.435 & 0.356 \\
		& & 0.607 & 1.222 & 0.602 & 0.740 & 0.656 \\
		\hline
		(mis, mis)& 800 & 0.000 & 0.258 & -0.062 & -0.183 & -0.016 \\
		& & 0.310 & 0.541 &  0.256 &  0.154 &  0.295 \\
		& & 0.607 & 0.807 &  0.555 &  0.475 &  0.581 \\[0.1cm]
		& 2000 & 0.000 & 0.262 & -0.056 & -0.168 & -0.003 \\
		& & 0.310 & 0.536 &  0.250 &  0.151 &  0.297 \\
		& & 0.607 & 0.792 &  0.543 &  0.467 &  0.584 \\
		\hline
	\end{tabular}
	\label{tab:simulation}
\end{table}

Besides this factor of model misspecification, we consider two sample sizes ($n=800$ or $2000$), and consider three values of $\gamma_T$ between 0 and 1. Also, we fix the values of the recall bias parameters as $(\eta_1, \eta_0) = (0.1, 0.1)$ throughout this simulation. Table~\ref{tab:simulation} shows the simulation results that are obtained from 2000 simulated datasets. The Crude method is essentially biased since confounding bias is not controlled. As shown in the table, Crude is biased in every scenario. Also, S$_{prop}$ is biased for all considered scenarios. The propensity score is estimated by regression $T_i^*$ on $(X_{i1}, X_{i2}, X_{i3}, X_{i4})$, thus the estimated score is biased. As we discussed in Section~\ref{ss:stratification}, we expect that when $\eta_0 = \eta_1$, stratification based on this biased score will adjust for confounding bias, but simulation shows that S$_{prop}$ always provides biased estimates. However, stratification based on prognostic scores, S$_{prog}$, shows great performance in recovering the true marginal COR. Especially for the case of (mis, mis), S$_{prog}$ shows the best performance. Finally, the ML method provides least biased estimates except for (mis, mis). Although it requires correct models for exposure and outcome, if either the exposure or outcome model is correctly specified, ML shows the best performance. However, when both models are misspecified, ML provides a more biased estimate than S$_{prog}$ does.

\section{Data Example: Child Abuse and Adult Anger}
\label{s:example}

We consider a retrospective case-control study to examine the question ``Does child abuse by either parent increase a likelihood toward to adult anger?'' This study is the 1993-94 sibling survey of the Wisconsin Longitudinal Study (WLS) that is publicly available. We define the exposure variable by combining two responses asking whether there was abuse in their childhood by father or mother respectively. The responses were measured in four categories: ``not at all'', ``a little'', ``some'', and ``a lot,'' and, by following \cite{springer2007long}, the exposure of child physical abuse is defined as an indicator of ``some'' or ``a lot'' to at least one of the two responses. Since these two questions were asked for adults to recall their childhood exposures, there may exist a systematic bias in recalling exposures. The outcome was initially measured as \cite{spielberger1988state}'s anger scale. We define a binary outcome variable; an individual is a case if his/her anger score is greater than or equal to 18 that is the 90th percentile of the measured anger scores. Also, seven covariates are considered: sex, age at the time of the interview, father's education, mother's education, parental income, farm background, and an indicator of parents' marital problems or single parent. See \cite{springer2007long, Small2013} for more details about the WLS data. 

With the same WLS data, \cite{springer2007long} used a logistic regression model and found that childhood physical abuse is associated with anger with odds ratio 2.02 with the 95\% CI (1.44, 2.84). However, in their discussion, they pointed out that the results may be affected by a tendency of under-reporting of abuse that is a common weakness in studies of child abuse and adult health. \cite{fergusson2000stability} summarized three properties in the nature of reporting/measuring exposure in childhood: (1) absence of false positive responses, (2) high rates (about 50\%) of false negative responses and (3) independence of psychiatric state and reporting errors. The first property means that if $\Tij = 0$, $\Tij^*$ cannot be 1. Second, recall bias is severe and there is a high proportion of reporting $\Tij^*=0$ when $\Tij = 1$ in reality. Finally, recall bias does not depends on the observed outcome, which implies $\zeta_0 = \zeta_1$. 

\begin{table}
	\centering
	\caption{Estimated marginal CORs for the effect of child abuse on adult anger without accounting for recall bias}
	\begin{tabular}{lccc}
		\hline
		Method & $\hat{\psi}$ & SE$(\log(\hat{\psi}))$ & 95\% CI \\
		\hline
		ML using logistic regressions (ML) & 1.84 & 0.16 & (1.36, 2.51)\\
		Stratification (S) & 1.79 & 0.16 & (1.31, 2.44)\\
		\hline
	\end{tabular}
	\label{tab:estimates}
\end{table}

We applied (i) the ML method by using logistic regression (ML) and (ii) stratification based on the prognostic score (S) for estimating the marginal odds ratio. For the ML method, the logistic outcome regression with the seven covariates without interaction terms is considered. For the S method, based on the same outcome model, the prognostic score can be estimated. Five strata are constructed by using the quintile values of the estimated prognostic score. 

\begin{table}
	\centering
	\caption{Sensitivity analysis of recall bias for five values of $\zeta_0= \zeta_1$. The estimates and 95\% confidence intervals are displayed for the maximum likelihood and stratification methods}
	\begin{tabular}{lcc}
		\hline
		& \multicolumn{2}{c}{Method}\\
		$(\zeta_0, \zeta_1)$ & ML & S \\
		\hline
		(.1,.1) & 1.87 (1.36, 2.57) & 1.81 (1.32, 2.49) \\
		(.2,.2) & 1.90 (1.36, 2.65) & 1.84 (1.33, 2.56) \\
		(.3,.3) & 1.94 (1.37, 2.75) & 1.89 (1.34, 2.65) \\
		(.4,.4) & 1.98 (1.37, 2.88) & 1.95 (1.36, 2.80) \\
		(.5,.5) & 2.03 (1.35, 3.06) & 2.06 (1.38, 3.06) \\
		\hline
	\end{tabular}
	\label{tab:sensi}
\end{table}

If there was no recall bias (i.e., $\zeta_0 = \zeta_1 = 0$ and $\Tij = \Tij^*$), the estimates are reported in Table~\ref{tab:estimates}. The first two methods for the marginal OR provide similar estimates, $\hat{\psi}_{ML} = 1.84$ with 95\% CI (1.35, 2.52) and $\hat{\psi}_{S} = 1.80$ with 95\% CI (1.30, 2.49). We conducted sensitivity analysis of recall bias with parameters $(\zeta_0, \zeta_1)$. Based on the previous literature, we focus on the line of $0 \leq \zeta_0 = \zeta_1 \leq 0.5$. The estimates for various values of $\zeta_0$ and $\zeta_1$ are shown in Table~\ref{tab:sensi}. All the estimates increase as $\zeta_0 = \zeta_1$ increases. Also, all the 95\% confidence intervals do not contain 1. This implies that the under-reporting issue does not alter the initial conclusion; on the contrary, it strengthens the conclusion that there is significant evidence that child abuse increases the odds of adult anger. Note that the variance of the estimates are obtained by using 500 bootstrapped samples.

\begin{figure}
	\centering
	\includegraphics[width = 160mm]{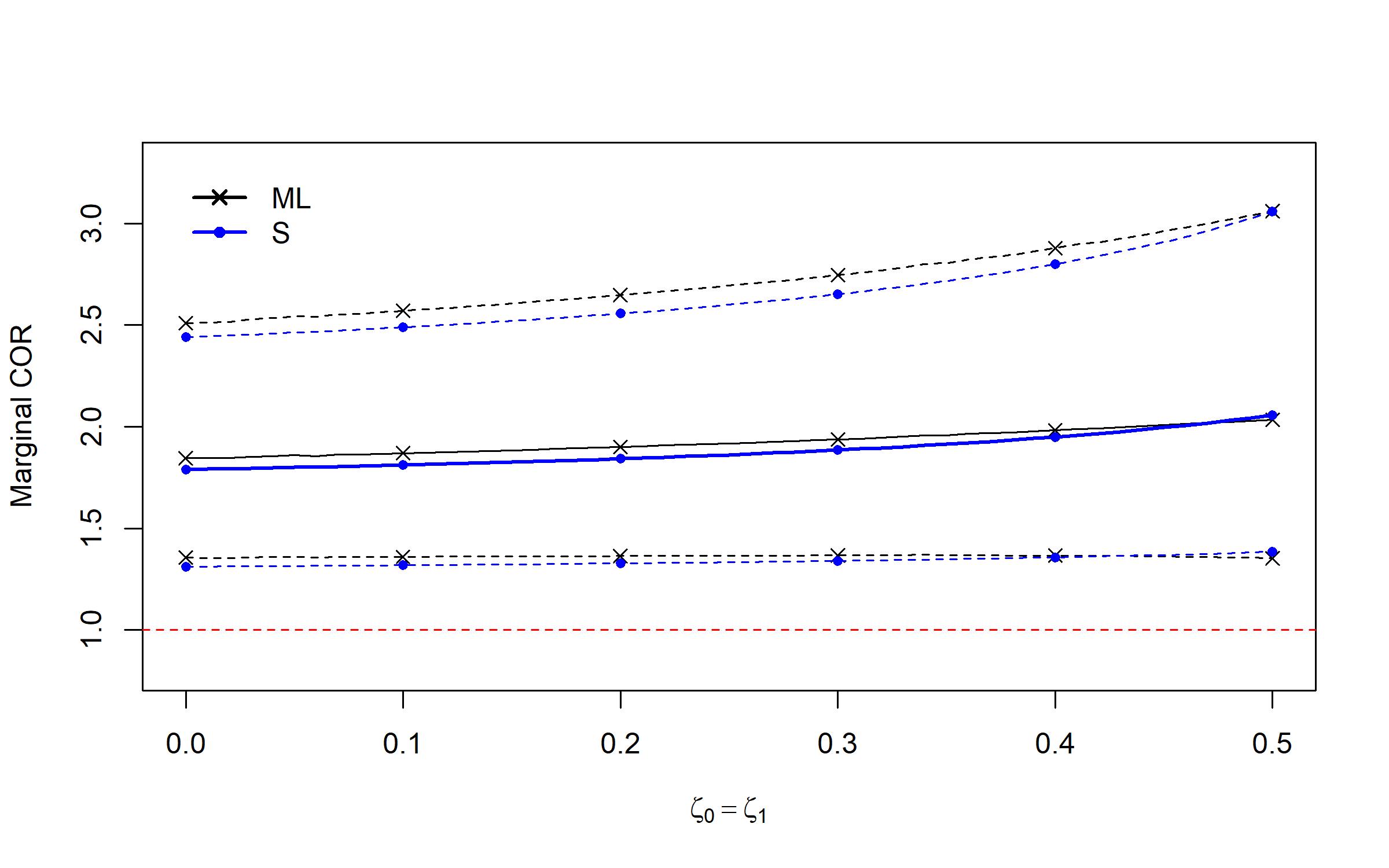}
	\caption{Estimates of the marginal OR with 95\% CIs across the line of $\zeta_0 = \zeta_1$}
	\label{fig:sensi}
\end{figure}

\begin{figure}
	\centering
	\includegraphics[width = 160mm]{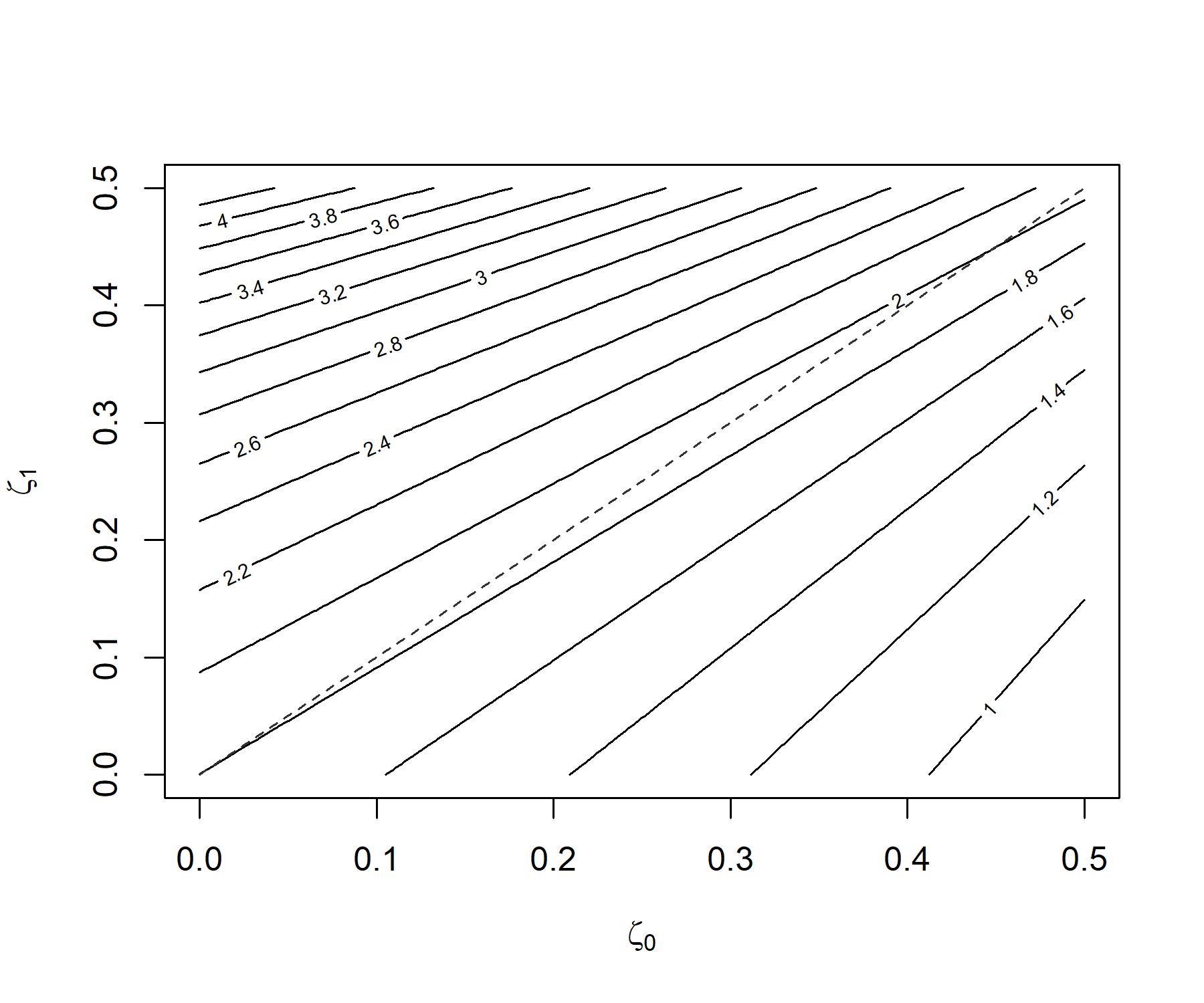}
	\caption{A contour plot for the values of $(\zeta_0, \zeta_1)$ in the region $0 \leq \zeta_0, \zeta_1 \leq 0.5$. Recovered estimates of $\psi$ are obtained from the stratification method. The diagonal line corresponds to the line of $\zeta_0 = \zeta_1$}
	\label{fig:sensi_contour}
\end{figure}

Although the ML method uses an additional model assumption that the S method does not, they provide almost identical results. For a larger value of $\zeta_0 = \zeta_1$, the estimates start to increase rapidly. The curves of the estimates across $\zeta_0 = \zeta_1$ are displayed in Figure~\ref{fig:sensi}. For all values of $\zeta_0 = \zeta_1$, the confidence bands are located far above the red line that indicates the null effect. We further extended the sensitivity analysis on the line of $\zeta_0 = \zeta_1$ to the region of $0 \leq \zeta_0, \zeta_1 \leq 0.5$. Since the ML and S methods provide similar results, we present the sensitivity analysis based on the S method. Figure~\ref{fig:sensi_contour} shows a contour plot of the estimated marginal ORs for the values of $\zeta_0$ and $\zeta_1$ in this region. The dashed diagonal line indicates the line of $\zeta_0 = \zeta_1$. The estimates along this line is represented in Figure~\ref{fig:sensi}. As shown in this figure, most of the estimates are above one. For the region of $\zeta_0 \geq \zeta_1 + 0.4$ for $0 \leq \zeta_1 \leq 0.1$, the estimates are below 1.

As discussed in Section~\ref{ss:rfactor}, the R-factor may be of interest. The initial conclusion obtained when $\zeta_0 = \zeta_1 = 0$ is that there is a strong effect of child abuse on adult anger. The R-factor can be considered when this conclusion can be altered. Since Proposition~\ref{prop3} shows that $\hat{\psi}_S(\zeta_0, \zeta_1)$ is a decreasing function of $\zeta_0$, we fix $\zeta_1=0$ and need to find the minimum value of $\zeta_0$ such that the 95\% CI contains 1. The R-factor is computed as 0.24 in this application. This shows that unless more than 24\% of controls under-reported their exposures while all cases correctly reported, the initial conclusion remains unchanged. Therefore, the R-factor can be used for examining the robustness to recall bias.   


\section{Discussion}
\label{sec:discussion}

In this paper, we have introduced a causal inference framework for case-control studies while accounting for recall bias, by using causal estimands such as marginal, conditional, and common CORs. We considered a set of two tuning parameters that characterizes all possible combinations of recall bias. We have proposed two estimation approaches for recovering the marginal COR (maximum likelihood and stratification) in the presence of recall bias. Our proposed approach has the following features. First, to best our knowledge, ours is the first attempt to estimate the marginal COR accounting for recall bias while adjusting for confounding bias. Second, we demonstrated theoretically and empirically that  failing to account for recall bias can lead to substantial bias in the estimation of the COR. We also showed the easy-to-check conditions when $\psi^*$ is greater than $\psi$ or vice versa. Third, we developed two estimation approaches, ML and stratification, based on the identification results in causal inference. In particular, stratification can be used to reduce the risk of model misspecification. Finally, we developed sensitivity analysis and introduced the R-factor which can provide information about how much recall bias is needed to qualitatively alter the causal conclusion. This will provide a practical guidance for practitioners to examine robustness of their findings.


We proposed and compared two estimation methods for recovering the marginal COR. However, both rely on model assumptions. Though the stratification method requires less than the ML method does, prognostic stratification requires an additional assumption that there is no effect modification. Another limitation is that the two tuning parameters may be too simple to illustrate recall bias. We assumed that the misclassified probabilities $(\eta_0, \eta_1)$ or $(\zeta_0, \zeta_1)$ are equal for all individuals, but this may be not realistic. Instead of assigning the same probability, we may restrict the probability within a certain interval. We are planning to extend the proposed method to matching that usually requires lesser model assumptions. We left this as our future research. 

Although our main goal is to make inference for the marginal COR, in sparse data cases such as matched case-control study data, methods for estimating the marginal COR may fail to provide a meaningful estimate. In such cases, the Mantel-Haenszel (MH) method can be used for summarizing a large number strata and estimating a common OR across strata. The MH estimator is not an appropriate estimator for the marginal COR since the common COR is typically not the same as the marginal COR \citep{austin2007performance}. However, the MH estimator can work well if the stratum-specific CORs are expected to be equal. Also, it can produce a simple summary of several contingency tables. We developed a new MH-type estimation method for a common COR that accounts for recall bias that is described in the Supplementary Material. The MH method uses matching to create strata, but does not require any of the model assumptions that the ML method assumes.  Furthermore, the MH method provides a closed form estimator so that the impact of recall bias can be analytically tractable. However, it requires an assumption for the target parameter that the stratum-specific CORs are identical. This may be implausible in some cases.

\section{Software}
\label{sec:software}

The methods described in this paper are available at \url{https://github.com/kwonsang/recall_bias_case_control_study}.

\section*{Supplementary Material}
\label{sec:supplementary}

Supplementary material is available online. These materials contain the Mantel-Haenszel method for estimating the common COR. Also, the population marginal COR is discussed. Proofs for Propositions 1,2,3 are described and the variance estimators are discussed.

\section*{Acknowledgments}

{\it Conflict of Interest}: The idea behind the development of this paper came from doing a consultation on a lawsuit against Colgate, but in a different context with a different data set.

\section*{Funding}

This work was supported by NIH grants (R01ES026217, R01MD012769, R01ES028033, 1R01ES030616, 1R01AG066793-01R01, 1R01ES029950, R01ES028033-S1), Alfred P. Sloan Foundation (G-2020-13946) and Vice Provost for Research at Harvard University (Climate Change Solutions Fund).

\bibliographystyle{APA}
\bibliography{recall}

\end{document}